\documentclass[sigconf]{acmart}

\AtBeginDocument{%
  \providecommand\BibTeX{{%
    \normalfont B\kern-0.5em{\scshape i\kern-0.25em b}\kern-0.8em\TeX}}}

\setcopyright{acmcopyright}
\copyrightyear{2023}
\acmYear{2023}
\acmDOI{XXXXXXX.XXXXXXX}

\usepackage{multirow}
\usepackage[normalem]{ulem}
\useunder{\uline}{\ul}{}
\usepackage{makecell}
\usepackage{graphicx}
\usepackage{subfigure}
\newtheorem{myDef}{Definition}
\usepackage[ruled]{algorithm2e}
\usepackage{appendix}

\acmConference[Conference acronym 'XX]{Conference acronym ’WWW}{ APRIL 30–MAY 4, 2023}{AUSTIN, TEXAS, USA}
\acmBooktitle{} 
\acmPrice{15.00}
\acmISBN{978-1-4503-XXXX-X/18/06}

\begin{document}

\title{Addressing the Extreme Cold-Start Problem in Group Recommendation}


\author{Linxin Guo}
\email{guolinxin@cqu.edu.cn}
\affiliation{%
  \institution{Chongqing University}
  \country{China}
}

\author{Yinghui Tao}
\email{taoyinghui@cqu.edu.cn}
\affiliation{%
  \institution{Chongqing University}
  \country{China}
}

\author{Min Gao}
\authornote{Min Gao is the corresponding author.}
\email{gaomin@cqu.edu.cn}
\affiliation{%
  \institution{Chongqing University}
  \country{China}
}

\author{Junliang Yu}
\email{jl.yu@uq.edu.au}
\affiliation{%
  \institution{The University of Queensland}
  \country{Australia}
}

\author{Liang Zhao}
\email{liangzhao@cqu.edu.cn}
\affiliation{%
  \institution{Chongqing University}
  \country{China}
}

\author{Wentao Li}
\email{Wentao.Li@uts.edu.au}
\affiliation{%
  \institution{University of Technology Sydney}
  \country{Australia}
}


\begin{abstract}

The task of recommending items to a group of users, a.k.a. group recommendation, is receiving increasing attention. However, the cold-start problem inherent in recommender systems is amplified in group recommendation because interaction data between groups and items are extremely scarce in practice. Most existing work exploits associations between groups and items to mitigate the data scarcity problem. However, existing approaches inevitably fail in extreme cold-start scenarios where associations between groups and items are lacking. For this reason, we design a group recommendation model for \underline{EX}reme cold-star\underline{T} in group \underline{RE}commendation (EXTRE) suitable for the extreme cold start scenario. The basic idea behind EXTRE is to use the limit theory of graph convolutional neural networks to establish implicit associations between groups and items, and the derivation of these associations does not require explicit interaction data, making it suitable for cold start scenarios. The training process of EXTRE depends on the newly defined and interpretable concepts of consistency and discrepancy, other than commonly used negative sampling with pairwise ranking, which can improve the performance of the group recommendation. Extensive experiments validate the efficacy of the proposed model EXTRE.

\end{abstract}





\maketitle

\section{Introduction}
\begin{figure}
    \centering
    \includegraphics[scale=0.3]{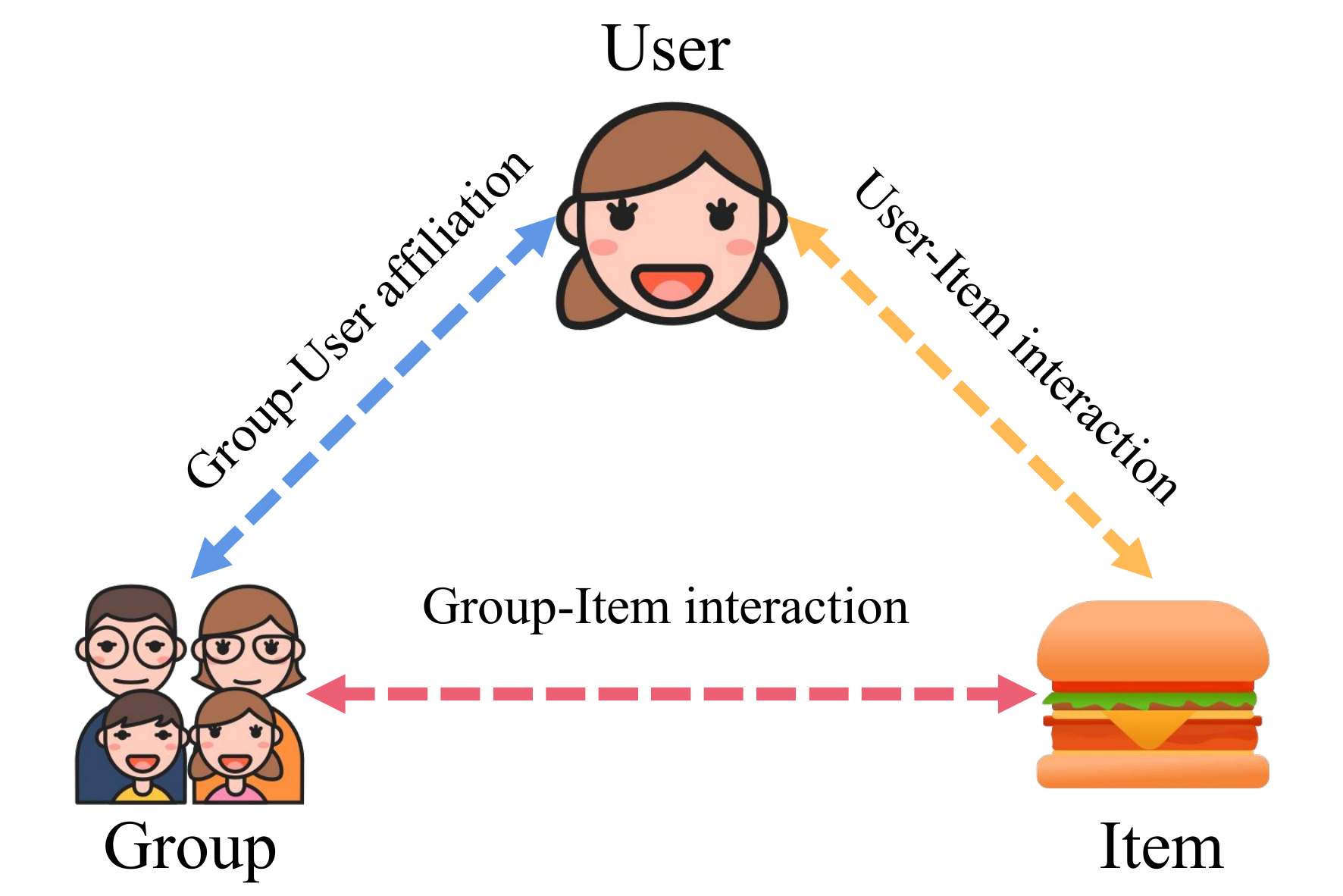}
    \caption{Group Recommendation}
    \label{fig:group-recommendation}

\end{figure}

Research on group recommendations \cite{amer2009group, ghaemmaghami2021deepgroup, pathak2017generating, cao2018attentive} has recently received increasing attention. Group recommendation aims to recommend an item to a group, such as recommending a travel route to a group of tourists. Group recommendation can deal with situations that general recommendations cannot. We can transfer group recommendation into a tripartite graph as shown in Fig.{\ref{fig:group-recommendation}}.

Group recommendation faces severe cold-start problems. Users in groups tend to interact with items much less frequently than individually. Therefore, group recommendation suffers from a more severe data sparsity problem. Especially in the cold-start phase, models have relative sufficient user-item interactions and group-user affiliations but severely lack group-item interactions.

Most of the existing research on the cold-start of group recommendation tries to alleviate the data sparsity problem by aggregating the representations of users. For example, Yin \emph{et al.}\cite{yin2019social} proposed a fusion of attention mechanism and bipartite graph embedding learning technology to simulate the social influence of each user and alleviate the data sparsity problem of group recommendation. Guo \emph{et al.} \cite{guo2020group} proposed a temporary group recommendation model that learns users' social influence based on a self-attention mechanism. It regards group decision-making as a multiple voting process and enriches the representation of groups through two aggregation methods. Cao \emph{et al.} \cite{cao2018attentive} proposed the group recommendation model AGREE, which introduces neural collaborative filtering and dynamically learned the fusion ratio of group embeddings through the attention mechanism. AGREE has a powerful recommendation effect for cold-start users without personal history interactions. Nevertheless, these models are still trained with group-item interactions to perform recommendations for cold-start group/item nodes. While in the extreme cold-start scenario, there is no group-item interactions available, the cold-start problem of group recommendation in such scenarios has not been solved.

The situation of no available group-item interactions also leads to invalidating negative sampling. Traditional negative sampling constructs triples that contain a group, an item the group has interacted with (positive sample), and an item the group has not interacted with (negative sample). Then the representations of the group and its positive sample will move close while the representations of the group and its negative sample go far away. However, we cannot decide if an item is a negative sample for a group since every group has no interactions with any items. Thus the negative sampling technique is unsuitable. 

To this end, we propose a model for \underline{EX}reme cold-star\underline{T} in group \underline{RE}commendation (EXTRE) to alleviate the extreme cold-start problem faced by group recommendation domains. Solving this extreme cold-start problem faces the following challenges: 
\begin{itemize}
\item How to devise a new method to substitute for negative sampling that can measure the relationship between nodes in the extreme cold-start scenario.
\item How to design a reasonable loss function to optimize the model while adapting the new node relationship measure method. 
\end{itemize}

To tackle the first challenge, we propose a method based on historical group-user and user-item relationships to extract the consistency and discrepancy between them. The consistency and discrepancy can directly measure the distance between a pair of nodes in extreme cold-start phase, while original positive and negative sampling cannot. Consistency and discrepancy can be calculated through three-node meta-paths in interaction graph. To handle the second challenge, we combine consistency and discrepancy with several effective loss functions and compare their outcomes, thus we can maximize the utilization of consistency and discrepancy and get more fine-grained representations of groups and items with consistency and discrepancy. The consistency and discrepancy can also be used to measure direct relationships in the tripartite interaction graph. When the group-item interactions become available, we can also utilize the consistency and discrepancy of group-item interactions to improve model performance.

The contributions of this paper can be summarized as follows:
\begin{itemize}
\item We propose a method addressing the extreme cold-start problem for group recommendation, which can achieve effective recommendation without group-item interactions in pre-training stage. Furthermore, the method can transition to scenarios with group-item interaction in fine-tuning stage.
\item We define consistency and discrepancy between special types of node pair (group-group, group-item, item-group, item-item) on meta-paths. They instruct how the node embeddings change during the training process. In extreme cold-start scenario, they perform much better than negative sampling.
\item We conduct extensive experiments on two public datasets to demonstrate the superiroity of our proposed EXTRE method in group and bundle recommendation domains, and comprehensively analyze the effectiveness of consistency and discrepancy and the benefits of pre-training and fine-tuning.

\end{itemize}

\section{Preliminary}
In this section, we state the symbols, related concepts, and the definition of group recommendation problems to be solved.
\begin{table}[]
\centering
\caption[]{Notations and explanations}
\label{tab:notation}
\scalebox{0.8}{
\begin{tabular}{c||l}
\hline
Notation & Explanation \\ \hline \hline
$\mathcal{G}$, $\mathcal{I}$, $\mathcal{U}$ & the sets of group, item, and user nodes\\
$g$, $i$, $u$ & a group / item / user node\\
($v_1$, $v_2$) & an arbitrary node pair\\
$\boldsymbol{Y}$, $\boldsymbol{X}$, $\boldsymbol{Z}$ & \begin{tabular}[c]{@{}l@{}} the matrix of group-item interaction, user-item interaction, \\and group-user affiliation \end{tabular}\\
$e_{g}$, $e_{i}$, $e_{u}$ & the embedding of group, item, and user nodes\\
$\mathcal{N}(g)$, $\mathcal{N}(i)$, $\mathcal{N}(u)$ & \begin{tabular}[c]{@{}l@{}} the set of neighbors of group node $g$, item node $i$, \\and user node $u$ \end{tabular}\\
$d_{v}$, $\delta_{v_1, v_2}$ & \begin{tabular}[c]{@{}l@{}} the degree of node $v$ and the composite term of the \\degrees of nodes $v_1$ and $v_2$\end{tabular}\\
$\alpha_{v_1v_2}$, $\beta_{v_1v_2}$ & the consistency and discrepancy from node $v_1$ to node $v_2$\\
$\mathcal{A}^{\Phi}$, $\mathcal{B}^{\Phi}$ & \begin{tabular}[c]{@{}l@{}} consistency and discrepancy sub-matrices extracted \\from meta-path $\Phi$ \end{tabular}\\
$\mathcal{A}_{group}$, $\mathcal{B}_{group}$ & \begin{tabular}[c]{@{}l@{}}consistency and discrepancy matrices extracted \\from group-item interactions  \end{tabular}\\
$\mathcal{A}_{user}$, $\mathcal{B}_{user}$ & \begin{tabular}[c]{@{}l@{}} consistency and discrepancy matrices extracted \\from user-item interactions and group-user affiliation\end{tabular}\\ \hline
\end{tabular}
}
\end{table}
\subsection{Relevant Concepts}
\begin{myDef}
\textbf{Tripartite graph} \cite{sun2013mining}. 
Given three different types of node sets $\mathcal{G}$, $\mathcal{U}$, and $\mathcal{I}$, a graph $G=(\mathcal{V},\mathcal{E}$) is called a tripartite graph, if $\mathcal{V} = \mathcal{G} \cup \mathcal{I} \cup \mathcal{U}$ and $\mathcal{E} \subseteq \mathcal{G} \times \mathcal{I} \cup \mathcal{U} \times \mathcal{I} \cup \mathcal{G} \times \mathcal{U}$. $G$ is also a heterogeneous graph.
\end{myDef}

\begin{myDef}
\textbf{Meta-path} \cite{sun2011pathsim}.
A meta-path $\Phi$ is defined on the tripartite graph $G=(\mathcal{V}, \mathcal{E})$, denoted as a path pattern in the form of $V_{1} V_{2} \cdots V_{l+1}$, which describes the composite semantic $R=R_{V_{1} V_{2}} \circ R_{V_{2} V_{3}} \circ \cdots \circ R_{V_{l} V_{l+1}}$ between the head node $V_{1}$ and the tail node $V_{l+1}$, where $\circ$ denotes the composition operator on semantics.
\end{myDef}

\subsection{Problem Statement}
Consider the set of group nodes $\mathcal{G} = \left \{ g_{1},\ldots,g_{|\mathcal{G}|}\right \}$, the set of item nodes $\mathcal{I} = \left \{ i_{1},\ldots,i_{|\mathcal{I}|}\right \}$, and the set of user nodes $\mathcal{U} = \left \{ u_{1},\ldots,u_{|\mathcal{U}|}\right \}$. $|\mathcal{G}|$, $|\mathcal{I}|$, and $|\mathcal{U}|$ represent the number of group, item, and user nodes, respectively. There are three node relationships among these three node types: group-item interaction, user-item interaction, and group-user affiliation. We use a sparse matrix $\boldsymbol{Y} \in\{0,1\}^{|\mathcal{G}| \times |\mathcal{I}|}$ to represent group interactions, describing implicit feedback between group nodes and item nodes. The element $y_{gi}$ = 1 indicates that group node $g$ has interacted with item node $i$, and $y_{gi}$ = 0 otherwise. Similarly, user interactions are represented by the sparse matrix $\boldsymbol{X} \in\{0,1\}^{|\mathcal{U}| \times |\mathcal{I}|}$, describing the implicit feedback of user nodes to item nodes. 
and the affiliation of the group and its users area denoted as $\boldsymbol{Z} \in\{0,1\}^{|\mathcal{G}| \times |\mathcal{U}|}$. 
The goal of group recommendation models is to provide a Top-K list of items for groups based on user historical interactions $\boldsymbol{X}$, group historical interactions $\boldsymbol{Y}$, and group-user affiliation $\boldsymbol{Z}$. Table \ref{tab:notation} summarizes the meaning of each symbol.

\textbf{Extreme cold-start of group recommendation}: Only user-item interaction $\boldsymbol{X}$ and group-user affiliation data $\boldsymbol{Z}$ is available at the extreme cold-start phase of group recommendation.

\section{Methodology}
In this section, we first introduce the overall structure of our model EXTRE. For relationships between group and item nodes, EXTRE uses the indirect ones to perform pre-training and then fine-tunes with the direct ones. To measure the indirect relationships, we derive two important concepts consistency and discrepancy from different types of meta-paths and explain how they are applied to the training process of EXTRE. At last, we propose a loss function derived from consistency and discrepancy to further improve the overall performance.
\begin{figure*}[htbp]

	\centering
	\scalebox{0.85}{
	\includegraphics[scale=0.35]{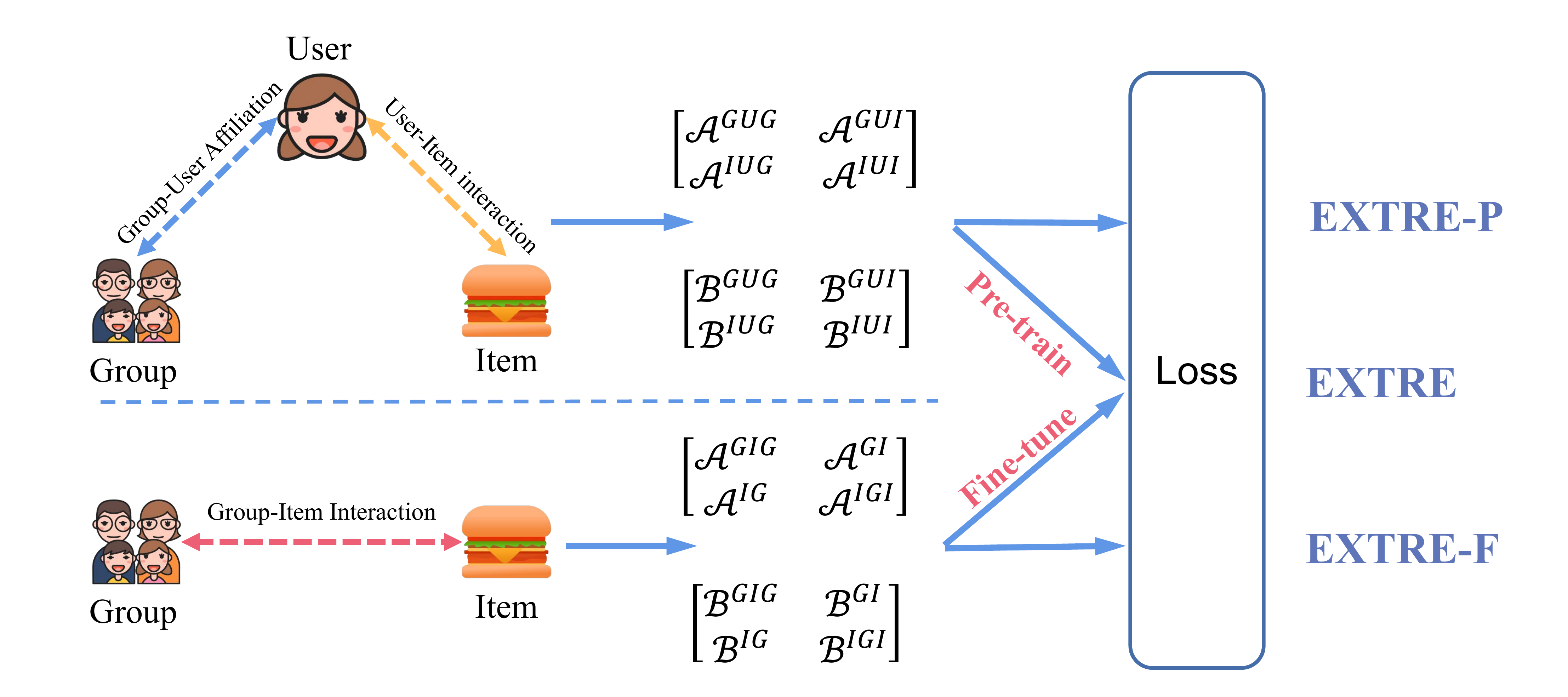}}
	\caption{The overall structure of EXTRE, the upper part shows the pre-training stage with group-user affiliations and user-item interactions, and the lower part indicates the fine-tuning stage with group-item interactions. EXTRE\_P and EXTRE\_F are two variants of EXTRE.}
	\label{fig:framework}

\end{figure*}

\subsection{Overall Process of EXTRE}
EXTRE adopts pre-training and fine-tuning technique like other methods \cite{yuan2020parameter,shin2021one4all,zhang2020general,sun2019bert4rec} to optimize the representation of groups and items, shown in Fig. \ref{fig:framework}. EXTRE performs pre-training using both user-item interactions and group-user affiliations, and performs fine-tuning if some group-item interactions are available. EXTRE derives consistencies and discrepancies from user-item and group-user relationships in pre-training stage, which is shows as two matrices in the upper part of Fig. \ref{fig:framework}. EXTRE derives consistencies and discrepancies from group-item relationships in fine-tuning stage, which is shows as two matrices in the lower part of Fig. \ref{fig:framework}. The four matrices consist of consistencies and discrepancies are conductive to extracting fine-grained node representations, which performs better than negative sampling. Finally, EXTRE uses consistency matrices and discrepancy matrices to construct loss functions and then update node representations. 

The pre-training and fine-tuning stages can work independently, forming two variants of EXTRE, i.e., EXTRE\_P and EXTRE\_F. EXTRE\_P is trained with only on user-item interactions and group-user affiliations, where P stands for pre-training. EXTRE\_F is trained with only on group-item interaction, where F stands for fine-tuning. Both of them can perform recommendations individually. Specifically, EXTRE\_P focuses on cold-start scenario, while EXTRE\_F works on warm-start context. Our complete recommendation model EXTRE combines EXTRE\_P and EXTRE\_F. 
The embedding matrices that need to be updated in EXTRE\_P and EXTRE\_F stages are $\mathbf{E}^{p}$ and $\mathbf{E}^{f}$. The pre-trained $\mathbf{E}^{p}$ is used to initialize $\mathbf{E}^{f}$ in the fine-tuning phase. In this way, a better initial value of embedding can be obtained, which reaches the convergence state faster. In addition, we preserve $\mathbf{E}^{p}$ as a fixed feature and concatenated $\mathbf{E}^{p}$ with $\mathbf{E}^{f}$ to form the final embedding $\mathbf{E} = \mathbf{E}^{f} \parallel \mathbf{E}^{p}$, where $\parallel$ is the concatenate operation. $\mathbf{E}$ is used for subsequent recommendation. 

Our model only updates $\mathbf{E}^{f}$ in the fine-tuning stage, which can avoid the node preferences extracted in the pre-training stage being erased by downstream tasks. That greatly preserves useful information in user interactions and affiliations. During the optimization process of EXTRE, two important concepts, i.e., consistency and discrepancy, are involved. The process of discovering them is introduced in the next subsection.

\subsection{Consistency and Discrepancy Based Fine-grained Representation Extraction from Complex Relationships}
In extreme cold-start scenario, there is no group-item interactions available, so that negative sampling cannot work, but there are far more complex relationships between nodes than the general recommendation. We propose a new method to learn the relationships between group and item nodes, based on consistency and discrepancy. In this section, we will explain these two terms' derivation and application.

\label{section:derivation}
The derivation of consistency and discrepancy starts from graph embedding learning task. Graph embedding learning task maps all the nodes in a graph into an embedding space and tries to preserve the information in this graph as much as possible. Recommender systems use these embeddings to perform recommendation. An institutional way to learning appropriate node embeddings is to fit all the edges in the graph. In the extreme cold-start scenario of group recommendation, we need to fit two kinds of relationships:

\begin{equation}
\label{equation:group-user relationship}
    \max \sum_{u\in \mathcal{N}(g)}e_g^{\top}e_u, \quad 
    \min \sum_{u\notin \mathcal{N}(g)}e_g^{\top}e_u,
\end{equation}

\begin{equation}
\label{equation:user-item relationship}
    \max \sum_{u\in \mathcal{N}(i)}e_i^{\top}e_u, \quad 
    \min \sum_{u\notin \mathcal{N}(i)}e_i^{\top}e_u.
\end{equation}

It is difficult to optimize these two kinds of edges simultaneously. Previous works always optimize only user and item embeddings (Eq.\ref{equation:user-item relationship}), and then use other techniques like attention\cite{vaswani2017attention} to aggregate users' embeddings into groups' ones. However, these aggregation methods always lack credibility, and it is not easy to estimate whether the aggregation process is reasonable. In our method, we optimize all these two kinds of edges together. Since LightGCN \cite{he2020lightgcn} has an excellent performance in mining structural information of recommender system graph, we learn user embeddings by LightGCN. The information transfer process of user embedding in the LightGCN network with self-connection can be formalized as

\begin{align}
\label{equation:LightGCN}
\begin{split}
e_u^{(l+1)} = \frac{1}{d_u+1}e_u^{(l)}
&+\sum_{i\in \mathcal{N}(u)}\frac{1}{\sqrt{(d_u+1)}\sqrt{(d_i+1)}}e_i^{(l)}\\
&+\sum_{g\in \mathcal{N}(u)}\frac{1}{\sqrt{(d_u+1)}\sqrt{(d_g+1)}}e_g^{(l)}.
\end{split}
\end{align}

The aggregation processes may consume a lot of time, and it's hard to decide how many aggregation processes are needed. Enlighted by UltraGCN\cite{mao2021ultragcn}, a convergent form of $e_u$ can omit these processes and reach a good representation. According to the aggregation formula of graph convolution, after passing through infinite layers of aggregation, the embedding of the model will converge to the following form: 

\begin{equation}
e_u=\lim _{l \rightarrow \infty} 
e_u^{(l+1)}=\lim _{l \rightarrow \infty} e_u^{(l)}.
\end{equation}

It can be seen that the node embedding is consistent before and after the convolution. Thus, Eq.\ref{equation:LightGCN} can be rewritten as:
\begin{equation}
\begin{split}
e_u = \frac{1}{d_u+1}e_u
&+\sum_{i\in \mathcal{N}(u)}\frac{1}{\sqrt{(d_u+1)}\sqrt{(d_i+1)}}e_i\\
&+\sum_{g\in \mathcal{N}(u)}\frac{1}{\sqrt{(d_u+1)}\sqrt{(d_g+1)}}e_g.
\end{split}
\end{equation}

After combining and simplifying, the following formula can be obtained:
\begin{equation}
\label{equation:converge_e_u}
e_u=\sum_{g \in \mathcal{N}(u)}
\frac{1}{d_u} \sqrt{\frac{d_u+1}{d_g+1}}e_g + 
\sum_{i \in \mathcal{N}\left(i\right)} 
\frac{1}{d_u} \sqrt{\frac{d_u+1}{d_i+1}}e_i. 
\end{equation}
Here we use $\delta_{u, g}$ and $\delta_{u, i}$ to represent the composite term of the degree of user $u$, group $g$, and item $i$:
\begin{equation}
\label{equation:delta}
\delta_{u,g} = \frac{1}{d_u}\sqrt{\frac{d_u+1}{d_g+1}},\quad
\delta_{u,i} = \frac{1}{d_u}\sqrt{\frac{d_u+1}{d_i+1}},
\end{equation}
and then we can simplify Eq.\ref{equation:converge_e_u} as the following form:
\begin{equation}
\label{equation:e_u}
e_u=\sum_{g \in \mathcal{N}\left(u\right)} \delta_{u, g} e_{g}+\sum_{i \in \mathcal{N}\left(u\right)} \delta_{u, i} e_{i}.
\end{equation}

Now we have the formula that uses group embeddings and item embeddings to represent user embeddings, i.e., Eq.\ref{equation:e_u}. To further utilize group-user relationships, we can substitute Eq.\ref{equation:e_u} into Eq.\ref{equation:group-user relationship}, the max part can be rewritten as:

\begin{equation}
\label{equation:group-user w/o user}
\begin{split}
\max \sum_{u \in \mathcal{N}(g)} e_g^{\top} e_u = 
\max (&\sum_{u \in \mathcal{N}(g)} \sum_{g^{\prime} \in \mathcal{N}\left(u\right)} \delta_{u, g^{\prime}} e_g^{\top} e_{g^{\prime}}\\
+&\sum_{u \in \mathcal{N}(g)} \sum_{i \in \mathcal{N}\left(u\right)} \delta_{u, i} e_{g}^{\top} e_{i}).
\end{split}
\end{equation}

This substitution eliminate the users' embedding $e_u$. So we can utilize the relationship of user nodes without training their embeddings, and then reduce the memory cost and computing time. 
To facilitate the optimization of the model, we convert Eq.\ref{equation:group-user w/o user} into the form of a loss function:

\begin{equation}
\label{equation:loss_gu_max}
\begin{split}
\mathcal{L}_{gu\_max }=&\sum_{\left(g, u, g^{\prime}\right) \in Q_{gug^\prime}} \delta_{u, g^{\prime}}\left(1-\sigma\left(e_{g}^{\top} e_{g^{\prime}}\right)\right)\\
+&\sum_{\left(g, u, i\right) \in Q_{gui}} \delta_{u, i}\left(1-\sigma\left(e_{g}^{\top} e_{i}\right)\right),
\end{split}
\end{equation}
where $\sigma(\cdot)$ is the sigmoid function, which maps the inner products result of node embeddings between 0 and 1. Sigmoid function can provide a larger gradient if its input is closed to 0, this can accelerate the training process. $Q_{gug}$ is the set of Group-User-Group meta-paths ($g,u,g^\prime$) composed of group node $g$ and its first-order user neighbors $u$ and second-order group neighbors $g^\prime$. Similarly, $Q_{gui}$ is the set of Group-User-Item meta-paths ($g, u, i$):
\begin{align}
&Q_{gug^\prime}=\left\{\left(g, u, g^{\prime}\right) \mid u \in \mathcal{N}(g), g^{\prime} \in \mathcal{N}\left(u\right)\right\}, \notag \\
&Q_{gui}=\left\{\left(g, u, i\right) \mid u \in \mathcal{N}(g), i \in \mathcal{N}\left(u\right)\right\}.
\end{align}

Similarly, we can get the loss from the min part of Eq.(\ref{equation:group-user relationship}):
\begin{equation}
\label{equation:loss_gu_min}
\begin{split}
\mathcal{L}_{gu\_min }=&\sum_{\tiny{\left(g, \tilde{u}, g^{\prime}\right)} \in Q_{g \tilde{u} g^\prime}} \delta_{\tilde{u}, g^{\prime}}\sigma\left(e_{g}^{\top} e_{g^{\prime}}\right)\\
+&\sum_{\tiny{\left(g, \tilde{u}, i\right) \in Q_{g \tilde{u} i}}} \delta_{\tilde{u}, i}\sigma\left(e_{g}^{\top} e_{i}\right). 
\end{split}
\end{equation}
The set $Q_{g \tilde{u} g^\prime}$ is different from $Q_{g u g^\prime}$. The triples $\left(g, \tilde{u}, g^{\prime}\right)$ in it represent that user $\tilde{u}$ is not the neighbor of group $g$. And group $g^{\prime}$ is the neighbor of user $\tilde{u}$. Here we expand the definition of meta-path to a generalized meta-path Group$\sim$User-Group, where $\sim$ indicates there is no edge. So is $Q_{g \tilde{u} i}$:
\begin{align}
&Q_{g \tilde{u} g^\prime}=\left\{\left(g, \tilde{u}, g^{\prime}\right) \mid \tilde{u} \notin \mathcal{N}(g), g^{ \prime} \in \mathcal{N}\left(\tilde{u}\right)\right\}, \notag \\
&Q_{g \tilde{u} i}=\left\{\left(g, \tilde{u}, i\right) \mid \tilde{u} \notin \mathcal{N}(g), i \in \mathcal{N}\left(\tilde{u}\right)\right\}.
\end{align}

Consider the loss functions $\mathcal{L}_{gu\_max}$ (Eq.\ref{equation:loss_gu_max}) and $\mathcal{L}_{gu\_min}$ (Eq.\ref{equation:loss_gu_min}). Their first terms are about meta-paths with group nodes $g$ and $g^\prime$ as the head node and tail node, and their second terms are about meta-paths with group $g$ as head node and item $i$ as the tail node. Thus, we can reorganize these loss functions according to the node types of meta-paths. We take their first terms as an example; given an arbitrary group node pair ($g, g^\prime$), its losses in $\mathcal{L}_{gu\_max}$ and $\mathcal{L}_{gu\_min}$ can be combined into the following form:

\begin{equation}
\begin{split}
l_{g g^{\prime}}=&\sum_{u \in \mathcal{N}(g), u \in \mathcal{N}\left(g^{\prime}\right)} \delta_{u, g^{\prime}}\left(1-\sigma\left(e_{g}^{\top} e_{g^{\prime}}\right)\right)\\
+&\sum_{\tilde{u}  \notin \mathcal{N}(g), \tilde{u} \in \mathcal{N}\left(g^{\prime}\right)} \delta_{\tilde{u}, g^{\prime}} \sigma\left(e_{g}^{\top} e_{g^{\prime}}\right).
\end{split}
\end{equation}

We omit the constant that independent to $e_g$ and $e_{g^\prime}$, and use $\alpha_{gg^\prime}$ and $\beta_{gg^\prime}$ to denote the constants that concern about $e_g$ and $e_{g^\prime}$:
\begin{equation}
    \alpha_{gg^\prime} = \sum_{u \in \mathcal{N}(g), u \in \mathcal{N}\left(g^{\prime}\right)} \delta_{u, g^{\prime}},\quad
    \beta_{gg^\prime} = \sum_{\tilde{u}  \notin \mathcal{N}(g), \tilde{u} \in \mathcal{N}\left(g^{\prime}\right)} \delta_{\tilde{u}, g^{\prime}},
\end{equation}
thus the loss function about a pair of group nodes $g$ and $g^\prime$ is:
\begin{equation}
    l_{gg^{\prime}} = (\beta_{gg^\prime} - \alpha_{gg^\prime})\sigma(e_g^\top e_{g^\prime}).
\end{equation}

Note that the loss function $l_{gg^\prime}$ depends on two similar meta-path set $Q_{gug^\prime}$ and $Q_{g\tilde{u}g^\prime}$, all the meta-paths in these two sets have same node types, the only different is the first edge in these meta-paths. By constructing similar meta-path sets, we can apply this method on other types of node pairs. For example, give an arbitrary node pair ($g,i$), we can construct two meta-path set $Q_{gui}$ and $Q_{g\tilde{u}i}$ and derive its loss function:
\begin{equation}
    l_{gi} = (\beta_{gi} - \alpha_{gi})\sigma(e_g^{\top}e_i).
\end{equation}
The node pairs can be group-group, group-item, item-group, and item-item. It is worthy noting that the two nodes in a pair is ordered due to the asymmetry of consistency and discrepancy. Finally, we can give out the loss function of a node pair ($v_1,v_2$) based on meta-paths with the user node as their middle node:
\begin{equation}
\label{equation:combine_loss}
   \mathcal{L}=\sum_{v_1, v_2 \in \mathcal{G\cup I}}\left(\beta_{v_1v_2}-\alpha_{v_1v_2}\right) \sigma\left(e_{v_1}^{\top} e_{v_2}\right).
\end{equation}

It is easy to discover that the coefficients $\alpha_{v_1v_2}$ and $\beta_{v_1v_2}$ decide the distance between the embeddings of two nodes. Specifically, a larger $\alpha_{v_1v_2}$ requires the two embeddings closer while a larger $\beta_{v_1v_2}$ forces the two embeddings farther away. Since we can calculate the two coefficients between any pair of nodes that can construct two meta-path sets, the two coefficients widely exist between nodes. We name the two coefficients $\alpha$ and $\beta$ as \textbf{consistency} and \textbf{discrepancy} separately; thus, the indirect relationship between group and item nodes is established.

\subsection{The training of EXTRE based on Consistency and Discrepancy}
An experiment (Fig.\ref{fig:loss}) shows that directly utilizing Eq.\ref{equation:combine_loss} to optimize node embeddings still leads to suboptimal result. The loss function Eq.\ref{equation:combine_loss} does not reduce the difficulty of optimization, the model still tries to fit two types of edges. But we can further utilize the consistency and discrepancy. Inspired by contrastive learning\cite{gutmann2010noise}, we re-design a loss function of the model:
\begin{equation}
\mathcal{L}=\sum_{(v_1, v_2) \in Q}-\log \frac{\alpha_{v_1 v_2} \exp \left(\cos \left(\mathbf{e}_{v_1}, \mathbf{e}_{v_2}\right) / \tau\right)}{\sum_{\tilde{v} \in \mathcal{G\cup I}} \beta_{v_1 \tilde{v}} \exp \left(\cos \left(\mathbf{e}_{v_1}, \mathbf{e}_{\tilde{v}}\right) / \tau\right)},
\label{equation:contrastive}
\end{equation}
where $Q=\left\{(v_1, v_2) \mid \alpha_{v_1v_2}>0\right\}$ is the set of node pairs whose consistency is positive. We use cosine similarity $\cos \left(\cdot, \cdot \right)$ for the similarity calculation between nodes. In Eq.\ref{equation:contrastive}, consistency and discrepancy are independent, during the optimization process, the consistency $\alpha_{v_1v_2}$ forces the embeddings of nodes $v_1$ and $v_2$ close and the discrepancy $\beta_{v_1 \tilde{v}}$ pushes away the embeddings of nodes $v_1$ and $\tilde{v}$. Moreover, the contrastive loss has a temperature hype-parameter $\tau$, which can adjust the focus of model training. Specifically, when the data noise is too large, a large temperature $\tau$ is applied and the model tends to update the embeddings of easily distinguished nodes and learn a rough preference. In contrast, when the quality of data obtained is relatively high, embeddings are accurately purified by reducing the temperature $\tau$ to distinguish similar nodes.

\subsection{The Generalized Form of Consistency and Discrepancy}
\label{section:generalization}
We derive the original consistency and discrepancy from eight types of meta-path, i.e., Group-User-Group ($GUG$), Group-User-Item ($GUI$), Item-User-Group ($IUG$), Item-User-Item ($IUI$) for consistency and Group$\sim$User-Group ($GUG$), Group$\sim$User-Item ($GUI$), Item$\sim$User-Group ($IUG$), Item$\sim$User-Item ($IUG$) for discrepancy. Since the four meta-path types for consistency and the four meta-path types for discrepancy are one-to-one correspondences, we use the same abbreviations for each meta-path type pair. These definitions are only suitable for extreme cold-start scenario. Since the user node type is in the middle of all the four types of meta-path, We use matrices with subscript \textit{user} to denote consistency and discrepancy over these eight types of meta-path:
\begin{equation}
\mathcal{A}_{user}=\left[\begin{array}{ll}
\mathcal{A}^{GUG} & \mathcal{A}^{GUI} \\
\mathcal{A}^{IUG} & \mathcal{A}^{IUI}
\end{array}\right], \quad 
\mathcal{B}_{user}=\left[\begin{array}{ll}
\mathcal{B}^{GUG} & \mathcal{B}^{GUI} \\
\mathcal{B}^{IUG} & \mathcal{B}^{IUI}
\end{array}\right],
\end{equation}
where $\mathcal{A}^{GUG}=\left[\alpha_{gg^{\prime}}\right]^{|\mathcal{G}| \times |\mathcal{G}|}$ represents the node consistency extracted through the meta-path set Group-User-Group ($GUG$). So are the other sub-matrices. In the pre-training stage, elements in these matrices are used.

We need similar matrices that can be used in the fine-tuning stage. In the fine-tuning stage, group-item interactions become accessible, so there are more affluent meta-paths can be used. Here we consider another four types of meta-path that do not contain user nodes, i.e., Group-Item-Group ($GIG$), Item-Group-Item ($IGI$) for consistency, Group$\sim$Item-Group ($GIG$), Item$\sim$Group-Item ($IGI$) for discrepancy. Since in the fine-tuning stage we can access to group-group interactions, we also consider the direct relationships between groups and items, no matter there is an edge between them. Following the asymmetry of consistency and discrepancy, we denote these direct relationships as ($TO$) and ($OT$). The similar matrices are denoted as follow:
\begin{equation}
\mathcal{A}_{group}=\left[\begin{array}{ll}
\mathcal{A}^{GIG} & \mathcal{A}^{GI} \\
\mathcal{A}^{IG} & \mathcal{A}^{IGI}
\end{array}\right], \quad 
\mathcal{B}_{group}=\left[\begin{array}{ll}
\mathcal{B}^{GIG} & \mathcal{B}^{GI} \\
\mathcal{B}^{IG} & \mathcal{B}^{IGI}
\end{array}\right],
\end{equation}
It should be noted that the one-hop meta-path $GI$ and $IG$ cannot extract the node relationship like the above. Hence, we need to change them to another form. Inspired by \cite{mao2021ultragcn}, we directly use $\delta$ as a generalization of consistency and discrepancy:
\begin{equation}
\mathcal{A}^{GI}=\left[\delta_{gi} * Y_{gi} \right]^{|\mathcal{G}| \times |\mathcal{I}|}, \quad
\mathcal{B}^{GI}=\left[\delta_{gi} \right]^{|\mathcal{G}| \times |\mathcal{I}|},
\end{equation}
where $\boldsymbol{Y} \in\{0,1\}^{|\mathcal{G}| \times |\mathcal{I}|}$ is the group-item interaction matrix. Consistency $\alpha_{gi}$ is calculated if there is an interaction between group $g$ and item $i$. Otherwise, the consistency $\alpha_{gi}$ is 0. The node relationship on the meta-path $IG$ is similar to $GI$. With the elements in matrices $\mathcal{A}_{group}$ and $\mathcal{B}_{group}$, we can further fine-tune our model to gain a better performance.


\section{Experiments}
In this section, we first introduce the specific settings of experiments. Next, our model does not limit to group recommendation. For other recommendation scenarios, if appropriate meta-path sets can be constructed, then our method can be applied. So we also demonstrate the superiority of our EXTRE model in various recommendation scenarios through extensive experiments on two public datasets. Furthermore, we explore the effect of different available data types on the model learning node preferences. We conduct ablation experiments to verify the efficiency of the extracted node relationships and loss function. Some supplementary experiments can be seen in Appendix.\ref{section:supplementary experiments}.
\begin{table}
\centering

\caption{The statistics of the datasets}
\scalebox{0.75}{
\begin{tabular}{|c|c|c|c|c|c|c|} 
\hline
\multirow{2}{*}{Mafengwo} & \#Group  & \#User & \#Item & \begin{tabular}[c]{@{}c@{}}\#Group-item\\interactions\end{tabular}  & \begin{tabular}[c]{@{}c@{}}\#User-item\\interactions\end{tabular} & \begin{tabular}[c]{@{}c@{}}\#Avg.users\\per group\end{tabular}   \\ 
\cline{2-7}
                          & 995      & 5,275  & 1,513  & 3,595                                                               & 39,761                                                            & 7.19                                                             \\ 
\hline
\multirow{2}{*}{Youshu}   & \#Bundle & \#Item & \#User & \begin{tabular}[c]{@{}c@{}}\#User-bundle\\interactions\end{tabular} & \begin{tabular}[c]{@{}c@{}}\#User-Item\\interactions\end{tabular} & \begin{tabular}[c]{@{}c@{}}\#Avg.items\\per bundle\end{tabular}  \\ 
\cline{2-7}
                          & 4,771    & 32,770 & 8,039  & 51,377                                                              & 138,515                                                           & 37.03                                                            \\ 
\hline
\end{tabular}}
\label{tab:datasets}
\end{table}

\subsection{Experiment Setup}

\textbf{Datasets.} Two public datasets are used in this work: group recommendation dataset Mafengwo and bundle recommendation dataset Youshu. The details are shown in Table \ref{tab:datasets}.

\textbf{Evaluation Metrics.}
To demonstrate the superiority of our model, we use several typical evaluation metrics to evaluate the accuracy of model recommendation, including F1-Score@K, Precision@K, Recall@K, and NDCG@K. These metrics have been widely used in previous studies \cite{he2020lightgcn, chang2021bundle, jia2021hypergraph}. We additionally use Entropy, Item\_Coverage, and Gini\_Index to judge the diversity of recommendation results\cite{wang2022survey}. The more different items are in the recommendation list, the higher these three metrics are.

\textbf{Implementation Details.}
This paper aims to address the cold-start problem in group recommendation. Therefore, we impose severe data constraints on models which employ group-item interactions. Specifically, we train these models with only 5\% group-user interactions. 20\% of group-user interactions are used to evaluate the final performance. The remaining data is used as the valid set. Unlike existing work \cite{chen2019matching, cao2018attentive, jia2021hypergraph} that randomly selects 100 negative samples for each test data, we perform a Top-K ranking on all items (bundles) in the datasets. In this way, more accurate model performance evaluation results can be obtained. The parameters of baselines refer to the settings given in their paper. For the fairness of the comparison, we unify the embedding dimension of all models to 64 dimensions. In addition, we set the patience of the early stopping method to 10, i.e., when the model fails to improve its performance for 10 consecutive epochs, the training is stopped. For our EXTRE, we randomly initialize the node embeddings, employ the Adam optimizer \cite{kingma2014adam} to update the model parameters, and set the learning rate to 0.001. For Mafengwo and Youshu datasets, we set the temperature hyper-parameter $\tau$ of the fine-tuning model to 1 and 0.3, respectively, and the pre-training model to 3.8 and 1, respectively. The meta-path sets adopted for each recommendation task are shown in Table \ref{tab:metapath}.

\begin{table*}[htbp]
\caption{Performance comparison of EXTRE and baselines on group and bundle recommendation datasets. Bold numbers represent optimal performance, and underlined numbers indicate suboptimal results.}
\label{tab:performance comparison}
\centering
\scalebox{0.8}{
\begin{tabular}{llllllll}
\hline \hline
Datasets                   & Method        & Recall@10 & Recall@20 & Recall@30 & NDCG@10 & NDCG@20 & NDCG@30 \\ \hline
\multirow{13}{*}{\makecell{Mafengwo \\ (Group)}} & \multicolumn{7}{c}{Group-item interaction}                                                       \\ \cline{2-8} 
                           & BPR           & 0.06040       & 0.11468        & 0.16820        & 0.03169        & 0.04603     & 0.05828      \\
                           & LINE          & 0.02433       & 0.03163        & 0.05657        & 0.01430        & 0.01636     & 0.02194      \\
                           & NGCF          & 0.03914       & 0.07951        & 0.11346        & 0.01592        & 0.02625     & 0.03343      \\
                           & LightGCN      & 0.07661       & 0.13838        & 0.19434        & 0.03854        & 0.05571     & 0.06821      \\
                           & BUIR          & {\ul 0.09694}       & {\ul 0.15092}        & {\ul 0.19281}        & {\ul 0.05928}        & {\ul 0.07396}     & {\ul 0.08386}      \\ 
                           & EXTRE\_F       & \textbf{0.19537}(+101.54\%)       & \textbf{0.28056}(+85.90\% )       & \textbf{0.32377}(+67.92\%)       & \textbf{0.12683}(+113.95\%)       & \textbf{0.14927}(+101.83\%)     & \textbf{0.15937}(+90.04\% )     \\ \cline{2-8} 
                           & \multicolumn{7}{c}{Group-item interaction/User-item interaction/Group-user affiliation}                        \\ \cline{2-8} 
                           & AGREE         & 0.03884       & 0.06850        & 0.09939        & 0.01860       & 0.02678     & 0.03381      \\
                           & HCR     & {\ul 0.05586}       & {\ul 0.08993}        & {\ul 0.12185}       & {\ul 0.02752}       & {\ul 0.03268}     & {\ul 0.03877}     \\ 
                           & EXTRE          & \textbf{0.26944}(+177.95\%)       & \textbf{0.38333}(+154.00\%)       & \textbf{0.43086}(+123.46\% )       & \textbf{0.14018}(+136.47\%)       & \textbf{0.17145}(+131.81\%)     & \textbf{0.18248}(+117.60\%)       \\ \hline
\multirow{13}{*}{\makecell{Youshu\\(Bundle)}}   & \multicolumn{7}{c}{User-bundle interaction}                                                       \\ \cline{2-8} 
                           & BPR           & 0.01883       & 0.03367        & 0.04420        & 0.01216        & 0.01699     & 0.02016      \\
                           & LINE          & 0.02335       & 0.03577        & 0.04668        & 0.01702        & 0.02077     & 0.02386      \\
                           & NGCF          & 0.02515       & 0.04786        & 0.06755        & 0.01669        & 0.02385     & 0.02978      \\
                           & LightGCN      & 0.03621       & 0.05734        & 0.07274        & 0.02692        & 0.03348     & 0.03785      \\ 
                           & BUIR          & {\ul 0.05082}       & {\ul 0.07853}        & {\ul 0.10336}        & {\ul 0.03629}        & {\ul 0.04441}     & {\ul 0.05103}      \\
                           & EXTRE\_F       & \textbf{0.06475}(+27.41\%)       & \textbf{0.09999}(+27.32\% )        & \textbf{0.12174}(+17.78\%)        & \textbf{0.03977}(+9.59\%)        & \textbf{0.05007}(+12.74\% )     & \textbf{0.05627}(+10.27\% )      \\ \cline{2-8} 
                           & \multicolumn{7}{c}{User-bundle interaction/User-item interaction/Bundle-item affiliation}                        \\ \cline{2-8} 
                           & DAM           & 0.01947       & 0.03164        & 0.04521        & 0.01342        & 0.01687     & 0.02135      \\
                           & BGCN          & {\ul 0.05434}       & {\ul 0.08476}        & {\ul 0.10903}       & {\ul 0.03579}       & {\ul 0.04553}     & {\ul 0.05280}     \\
                           & EXTRE          & \textbf{0.09026}(+66.10\%)       & \textbf{0.12994}(+53.30\%)       & \textbf{0.15202}(+39.43\%)       & \textbf{0.06273}(+72.86\%)       & \textbf{0.07444}(+63.50\%)     & \textbf{0.08091}(+53.24\%)     \\\hline \hline
\end{tabular}}
\end{table*}

\subsection{Performance Comparison in Cold-Start Scenario}
Table \ref{tab:performance comparison} lists the performance of our EXTRE and baselines. In this experiment, we set K to 10, 20, and 30 for Recall@K and NDCG@K to evaluate all models. From Table \ref{tab:performance comparison}, we can draw the following observations and conclusions:

First, our model outperforms SOTA related models in group and bundle recommendations. In particular, the average improvement on the Mafengwo dataset is as high as 140.22\%, and the Youshu dataset also have an increase of 58.07\% compared to the best-performing baseline. Second, this scenario can be regarded as a general recommendation when only using group-item or user-bundle interactions. The variant EXTRE\_F is also highly competitive in the general recommendation domain. 

This result shows that it is feasible to transfer node relationships in EXTRE to other domains by replacing meta-paths. Although the consistency and discrepancy are derived from the group recommendation, they also exist in other scenarios. Therefore, we can refine the node preference in the general recommendation domain through consistency and discrepancy. The results validate our model's effectiveness for broad scenarios.

In addition, AGREE and HCR in group recommendation and DAM in bundle recommendation work exceptionally poorly. The possible reason is that the multi-task joint optimization method is unsuitable for the cold-start stage of group recommendation. In group recommendation, group-item interactions is less than user-item interactions. To better simulate the cold-start scenario, we only keep 5\% of group-item interactions. It leads to a severe imbalance of the number of user-item interactions versus group-item interactions. It should be noted that this setting is realistic because the number of user-item interactions in the actual scene is far more than group-item interactions. This difference leads to models which jointly optimize the group and user recommendations, unable to balance these two tasks, thus affecting the final recommendation effect. However, our EXTRE adopts pre-train\&fine-tune to mine these two kinds of interactions. It avoids the dilemma of multi-task joint optimization in the cold-start scenario of group recommendation.

The model BGCN does not take user-item interactions as a prediction task. Instead, it applies the graph convolution, which can also effectively avoid the data imbalance problem. BGCN is the only bundle recommendation model to maintain excellent performance after reducing bundle data. It also verifies our above conjecture again.

We believe that the superior performance of our EXTRE is mainly due to the following reasons: 1) The consistency and discrepancy we quantified can effectively guide the model to learn fine-grained node preferences. 2) Adopting the pre-train\&fine-tune mode improves the generality of our self-supervised task while effectively avoiding the dilemma of data imbalance between group-item and user-item interactions. 3) The loss function we adopted can significantly exploit the potential of these quantified node relationships.
\begin{table}[]
\caption{Performance comparison of EXTRE\_P and baselines.}
\label{tab:cold-start}
\centering
\scalebox{0.8}{
\begin{tabular}{c|l|llll}
\hline
\multicolumn{1}{l|}{Dataset}                                                 & Method   & Recall@10        & Recall@20        & NDCG@10          & NDCG@20          \\ \hline
\multirow{6}{*}{\begin{tabular}[c]{@{}c@{}}Mafengwo \\ (Group)\end{tabular}} & BPR      & 0.0604           & 0.11468          & 0.03169          & 0.04603          \\
                                                                             & LINE     & 0.02433          & 0.03163          & 0.0143           & 0.01636          \\
                                                                             & NGCF     & 0.03914          & 0.07951          & 0.01592          & 0.02625          \\
                                                                             & LightGCN & 0.07661          & 0.13838          & 0.03854          & 0.05571          \\
                                                                             & BUIR     & {\ul 0.09694}    & {\ul 0.15092}    & {\ul 0.05928}    & {\ul 0.07396}    \\ \cline{2-6} 
                                                                             & EXTRE\_P & \textbf{0.12747} & \textbf{0.16327} & \textbf{0.06534} & \textbf{0.07445} \\ \hline
\multirow{6}{*}{\begin{tabular}[c]{@{}c@{}}Youshu\\ (Bundle)\end{tabular}}   & BPR      & 0.01883          & 0.03367          & 0.01216          & 0.01699          \\
                                                                             & LINE     & 0.02335          & 0.03577          & 0.01702          & 0.02077          \\
                                                                             & NGCF     & 0.02515          & 0.04786          & 0.01669          & 0.02385          \\
                                                                             & LightGCN & {\ul 0.03621}    & {\ul 0.05734}    & {\ul 0.02692}    & {\ul 0.03348}    \\
                                                                             & BUIR     & \textbf{0.05082} & \textbf{0.07853} & \textbf{0.03629} & \textbf{0.04441} \\ \cline{2-6} 
                                                                             & EXTRE\_P & 0.02801          & 0.04762          & 0.01898          & 0.02506          \\ \hline
\end{tabular}}
\vspace{-0.4cm}
\end{table}

\subsection{Performance Comparison in Extreme Cold-Start Scenario}
In this subsection, we validate the recommendation performance of our model in extreme cold-start scenarios, i.e., training models without any group-item interactions. A far as we know, EXTRE is the first model that can completely abandon group-item interactions. No baseline can be compared with it in the same conditions. Therefore, we compare EXTRE\_P with models using user-item interactions. It can be seen from Table \ref{tab:cold-start} that our model can achieve the best performance on the Mafengwo dataset. EXTRE\_P is also second only to LightGCN and BUIR on the Youshu dataset. Achieving comparable results to supervised models without using group-item interactions is incredibly difficult. This undoubtedly verifies the adaptability of our model to extreme scenarios.

We believe that the reason is that the extracted supervision signals are multi-angle, which takes into account the consistency and discrepancy between nodes. Second, user-item interactions and group-user affiliations are larger than group-item interactions, so valuable information is also richer.

\begin{table}[]
\caption{Diversity contrast experiment}
\label{tab:Diversity}
\centering
\scalebox{0.75}{
\begin{tabular}{c|cccc}
\hline
Dataset                                                                     & \multicolumn{1}{c|}{Method}   & Entropy          & Item\_Coverage   & Gini\_Index      \\ \hline
\multirow{4}{*}{\begin{tabular}[c]{@{}c@{}}Mafengwo\\ (Group)\end{tabular}} & \multicolumn{1}{c|}{BUIR}     & 4.05071          & 0.20408          & 0.92057          \\ \cline{2-5} 
                                                                            & \multicolumn{1}{c|}{EXTRE\_P} & \textbf{4.66926} & \textbf{0.36549} & \textbf{0.84857} \\
                                                                            & \multicolumn{1}{c|}{EXTRE\_F} & 3.55253          & 0.19666          & 0.94717          \\
                                                                            & \multicolumn{1}{c|}{EXTRE}    & {\ul 4.22368}    & {\ul 0.25603}    & {\ul 0.90513}    \\ \hline
\multirow{4}{*}{\begin{tabular}[c]{@{}c@{}}Youshu\\ (Bundle)\end{tabular}}  & \multicolumn{1}{c|}{BGCN}     & 4.78748          & 0.17606          & 0.97487          \\ \cline{2-5} 
                                                                            & \multicolumn{1}{c|}{EXTRE\_P} & \textbf{6.66109} & \textbf{0.4366}  & \textbf{0.87105} \\
                                                                            & \multicolumn{1}{c|}{EXTRE\_F} & 4.75747          & 0.21924          & 0.95673          \\
                                                                            & \multicolumn{1}{c|}{EXTRE}    & {\ul 5.29849}    & {\ul 0.31524}    & {\ul 0.95475}    \\ \hline
\end{tabular}}
\end{table}

\subsection{Diversity Analysis}
In recommendation systems, excessive pursuit of accuracy will make the model inclined to exploit the known interests of users and lead to information cocoons. An good recommendation method needs to recommend accurately and cover more items as possible. Therefore, we verify the recommendation diversity of EXTRE and its variants by comparing the best-performing BUIR and BGCN models. Unlike Entropy and Item\_Coverage, the lower the value of Gini\_Index, the better the variety. From Table \ref{tab:performance comparison}, Table \ref{tab:Diversity}, and Table \ref{tab:Diversity}, it can be found that our model is better than those two baselines in terms of recommendation accuracy and diversity. Accuracy and diversity tend to trade-off like a seesaw, and improving one metric can affect the performance of the other. Therefore, it is challenging to maintain excellent performance while increasing diversity. We attribute this to pre-training task. 

As shown in Table \ref{tab:Diversity}, the performance of the fine-tuning model in diversity is not outstanding. Nevertheless, it is greatly improved by combining with the pre-training stage. To further verify our conjecture, we also show the performance of the pre-training model EXTRE\_P. Its performance in diversity is better than all other models due to the pre-training phase utilizing numerous user-item interactions and group-user affiliations for training. EXTRE\_P effectively transfers user preference to group preference, which in turn improves the diversity of the final model EXTRE.

From another perspective, roughly preferences learned through auxiliary data can enhance the diversity of recommendations while fine-grained preferences learned from group-item interactions are helpful for recommendation accuracy. Our whole model effectively incorporates the advantages of these data through pre-train\&fine-tune mode.

\subsection{Ablation Study}
In this subsection, we conduct ablation experiments on various components of the EXTRE model to verify the rationality of the EXTRE design.

EXTRE\_P and EXTRE\_F are pre-training and fine-tuning models, respectively. EXTRE\_R represents the model that exchanges pre-training and fine-tuning data, which utilizes group-item interactions for pre-training, and user-item interactions and group-user affiliation for fine-tuning. According to Table \ref{tab:Ablation}, it can be seen that whether the pre-training or fine-tuning stage is eliminated, the performance of the model will be reduced. It verifies that the pre-train\&fine-tune mode we adopted is effective. In addition, after swapping the pre-training and fine-tuned data, the model is also affected to varying degrees. After fine-tuning through user-item interactions and group-user affiliations, the performance is degraded on the Youshu dataset. It shows that group-item interaction data is more critical than user-item interactions for group recommendation. Therefore, if some group-item interactions are available, the use of them in the final stage can grasp the trend of models.

\begin{table}[]
\caption{performance of variants of EXTRE}
\label{tab:Ablation}
\centering
\scalebox{0.75}{
\begin{tabular}{c|c|cccc}
\hline
Dataset                                                                     & Method          & F1-Score         & Precision        & Recall           & NDCG             \\ \hline
\multirow{7}{*}{\begin{tabular}[c]{@{}c@{}}Mafengwo\\ (Group)\end{tabular}} & EXTRE\_P        & 0.01950          & 0.01037          & 0.16327          & 0.07445          \\
                                                                            & EXTRE\_F        & 0.03360          & 0.01787          & 0.28056          & 0.14927          \\
                                                                            & EXTRE\_R        & 0.04247          & 0.02259          & 0.35309          & \textbf{0.17474} \\ \cline{2-6} 
                                                                            & w/o alpha       & 0.02955          & 0.01574          & 0.24105          & 0.1164           \\
                                                                            & w/o beta        & 0.03490          & 0.01861          & 0.27901          & 0.11929          \\
                                                                            & w/o alpha\&beta & 0.02976          & 0.01583          & 0.24660          & 0.11773          \\ \cline{2-6} 
                                                                            & EXTRE           & \textbf{0.04743} & \textbf{0.02528} & \textbf{0.38333} & 0.17145          \\ \hline
\multirow{7}{*}{\begin{tabular}[c]{@{}c@{}}Youshu\\ (Bundle)\end{tabular}}  & EXTRE\_P        & 0.01351          & 0.00787          & 0.04762          & 0.02506          \\
                                                                            & EXTRE\_F        & 0.02074          & 0.01157          & 0.09999          & 0.05007          \\
                                                                            & EXTRE\_R        & 0.01705          & 0.00988          & 0.06205          & 0.03481          \\ \cline{2-6} 
                                                                            & w/o alpha       & 0.02771          & 0.01572          & 0.11656          & 0.06483          \\
                                                                            & w/o beta        & 0.01794          & 0.01139          & 0.04226          & 0.02798          \\
                                                                            & w/o alpha\&beta & 0.01798          & 0.01139          & 0.04271          & 0.02808          \\ \cline{2-6} 
                                                                            & EXTRE           & \textbf{0.03336} & \textbf{0.01913} & \textbf{0.12994} & \textbf{0.07444} \\ \hline
\end{tabular}}
\end{table}

The quantified consistency and discrepancy are fine-grained, reflecting the complex relationship between nodes better than the binary labels (0 and 1). This characteristic is also embodied in real life. To verify that the derived consistency and discrepancy are adequate, we also conducted ablation experiments on these relationships. Specifically, we replace the consistency $\alpha$ or discrepancy $\beta$ greater than 0 with 1, that is, to express these relationships in a coarse-grained way. It can be observed from Table \ref{tab:Ablation} that erasing the consistency or the discrepancy will reduce the recommendation performance of the model. It verifies that our quantification of consistency and discrepancy is valid. 

\begin{figure}[htbp]
	\centering
	\scalebox{0.4}{
	\includegraphics[width=15cm]{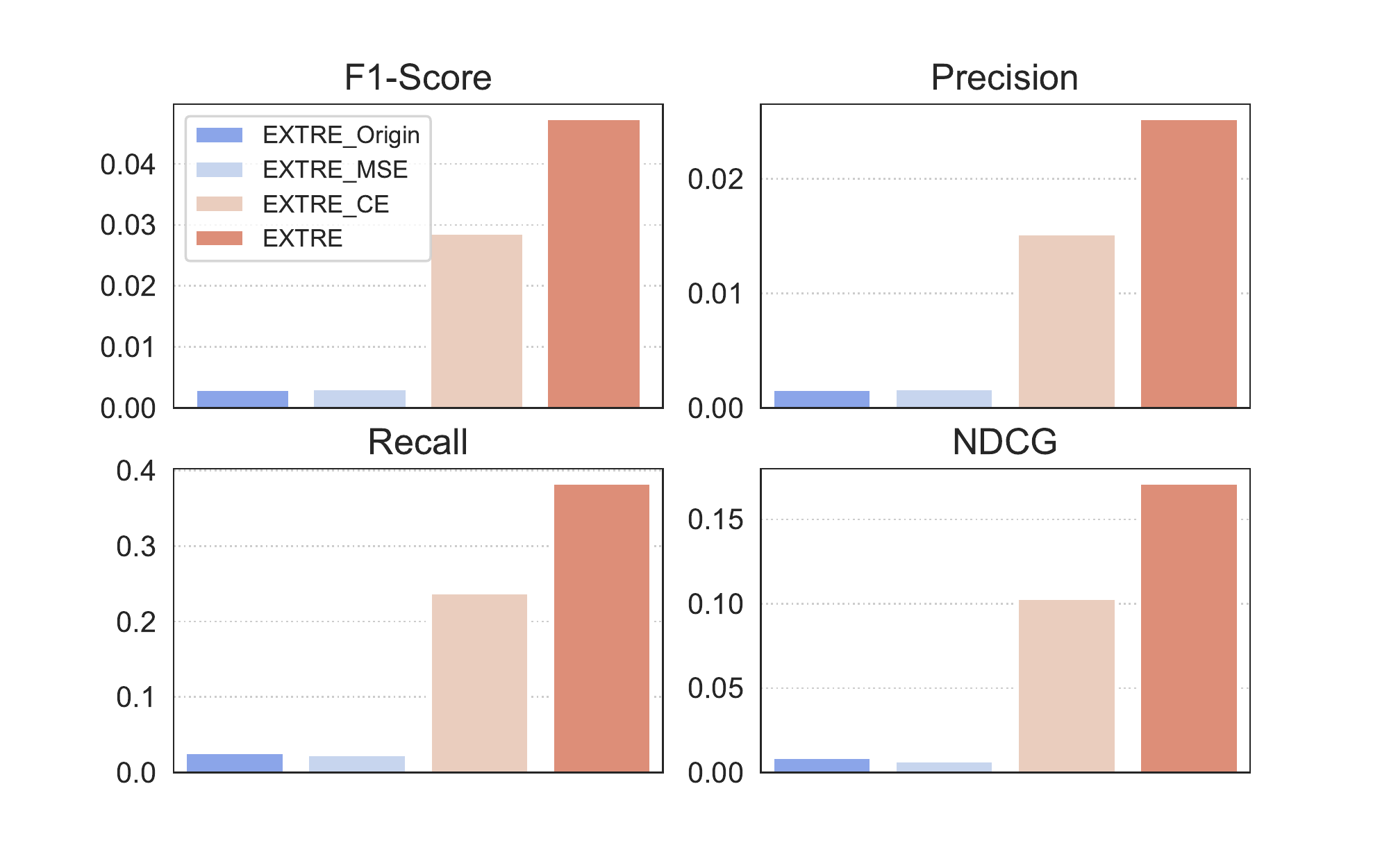}}
	\caption{Performance comparison of three different loss function of EXTRE on the Mafengwo dataset.}
	\label{fig:loss}

\end{figure}

Furthermore, to justify the design of our loss function, we replace the loss function of the EXTRE model. EXTRE\_Origin represents the originally derived loss (Eq.\ref{equation:combine_loss}). EXTRE\_MSE describes fitting $(\beta_{v_1v_2}-\alpha_{v_1v_2})$ using mean squared error \cite{hu2008collaborative}. EXTRE\_CE means that the model is trained using the cross-entropy loss \cite{he2017neural}. As can be seen from Fig. \ref{fig:loss}, EXTRE\_Origin and EXTRE\_MSE are incredibly ineffective. We think this is because the term $\alpha-\beta$ will erase the consistency and discrepancy. The performance of EXTRE\_CE and EXTRE, which use these node relationships separately, is better than EXTRE\_Origin and EXTRE\_MSE. The EXTRE with contrastive loss performed best due to its characteristics to distinguish relatively easy or difficult nodes by adjusting the temperature $\tau$. 

\section{Related Work}
The cold-start problem widely exists in recommendation scenarios. Lika \emph{et al.} summarized cold-start problem in recommendation into three types, i.e., recommend new items to users, recommend items to novel users, and recommend new items to novel users. \cite{lika2014facing} Schein \emph{et al.} \cite{schein2002methods} combined content and interactions to alleviate the data sparsity and provided metrics to measure the performance of recommedation. Lam \emph{et al.} \cite{lam2008addressing} developed a hybird method that recommends items to novel users. Jia \emph{et al.} \cite{jia2021hypergraph} utilized the user and group networks to form a two-channel hypergraph convolutional model HCR and modeled user and group preferences through multi-task joint optimization. Cai \emph{et al.}\cite{cai2022user} utilized the relational information to alleviate user attributes sparsity. Lara \emph{et al.}\cite{quijano2012case} supposed the novel user in a group is similar to other users in the same group. Ling \emph{et al.} \cite{yanxiang2013user} applied cluster algorithms on all entities and assumed the members of a cluster share some similar features. He \emph{et al.}\cite{he2020game} aimed to provide items to occasional groups by aggregating multi-view user representations with attention. On contrast, our model utilizes more fine-grained relationships in tripartite graphs and adopts corresponding loss function such that it performs much better in the cold-start scenario.

Pre-training and fine-tuning are used in various tasks.The critical parts of pre-training are how to generate augmented data and how to design a pretext task. Inspired by NLP pre-training model BERT \cite{devlin2018bert}, Chen \emph{et al.} \cite{chen2019bert4sessrec} and Zhang \emph{et al.} \cite{zhang2021unbert} design similar models for recommendation. And in recommendation scenarios, there are some side informations about users and items, Moreira \emph{et al.} \cite{moreira2021transformers} and Meng \emph{et al.} \cite{meng2021graph} propose pre-training models using these side informations. To utilize knowledge from other domains, Wang \emph{et al.} \cite{wang2021pre} pre-train model on one domain and then transfer the model to another domain to gain a better recommendation performance. 

\section{Conclusion and Future Work}
In this work, we propose a novel solution to the cold-start problem of group recommendation. Specifically, we deduce the formula from the group recommendation cold-start scenario's available data to extract the consistency and discrepancy between nodes. Afterward, we analyze and discuss these two relationships in detail and employ them as the self-generated supervision signal to guide model training. Through extensive experiments in various recommendation domains, the superiority and generality of our model are verified. In the future, we will combine consistency and discrepancy with advanced encoders to further develop and utilize these two node relationships.

\bibliographystyle{ACM-Reference-Format}
\bibliography{sample-base}

\clearpage

\appendix
\appendixpage
\section{the Asymmetry of Consistency and Discrepancy}
\label{section:asymmetry}
We need to calculate consistency and discrepancy over all node pairs with different order due to their asymmetry. Given an arbitrary node pair ($v_1,v_2$), the asymmetry of consistency and discrepancy means that $\alpha_{v_1v_2} \ne \alpha_{v_2v_1}$ and $\beta_{v_1v_2} \ne \beta_{v_2v_1}$. It is easy to prove that consistency and discrepancy are asymmetric. We use consistency as an example:
\begin{gather}
    \label{equation:asymmetry}
    \alpha_{v_1v_2} = \sum_{u \in \mathcal{N}(v_1),u \in \mathcal{N}(v_2)} \delta_{u,v_2}, \quad
    \alpha_{v_2v_1} = \sum_{u \in \mathcal{N}(v_2),u \in \mathcal{N}(v_1)} \delta_{u,v_1}.
\end{gather}

Eq.\ref{equation:asymmetry} shows that $\alpha_{v_1v_2}$ and $\alpha_{v_2v_1}$ have the same number of terms. But we cannot ensure every corresponding terms equal, i.e., $\delta_{u,v_2} = \delta_{u,v_1}$, due to the different degrees of $v_1$ and $v_2$. Similarly, discrepancy is asymmetric too.

\section{Algorithm}
The overall process of EXTRE is presented in Algorithm \ref{alg:algorithm}. 
\begin{algorithm}[h]

	\caption{The overall process of EXTRE}
	\LinesNumbered 
	\label{alg:algorithm}
	\KwIn{Group-item interaction matrix $Y$ \\
	\qquad \quad User-item interaction matrix $X$ \\
	\qquad \quad Group-user affiliation matrix $Z$ }
	\KwOut{Final node embeddings $\mathbf{E}$}
	\DontPrintSemicolon
	Extract $\mathcal{A}_{user}$ and $\mathcal{B}_{user}$ from user-item interaction $X$ and group-user affiliation $Z$\\
	Randomly initialize embedding $\mathbf{E}^{p}$\\
	\While{not converge}{
	    \For{each epoch}{
			Evaluate $\mathcal{L}$ through $\mathcal{A}_{user}$, $\mathcal{B}_{user}$ and $\mathbf{E}^{p}$ (Eq.\ref{equation:contrastive})\\
    		Back propagation and update embeddings $\mathbf{E}^{p}$\\
	    }
	}
	Extract $\mathcal{A}_{group}$ and $\mathcal{B}_{group}$ from group-item interaction $Y$\\
	Stop gradient of $\mathbf{E}^{p}$\\
	Initialize node embedding $\mathbf{E}^{f}$ with $\mathbf{E}^{p}$\\
	Concatenate final embeddings $\mathbf{E} \gets \mathbf{E}^{f} \parallel \mathbf{E}^{p}$ \\
	\While{not converge}{
	    \For{each epoch}{
			Evaluate $\mathcal{L}$ through $\mathcal{A}_{group}$, $\mathcal{B}_{group}$ and $\mathbf{E}$ (Eq.\ref{equation:contrastive})\\
    		Back propagation and update node embeddings $\mathbf{E}^{f}$\\
	    }
	}
    \textbf{return} Final node embeddings $\mathbf{E}$\\
\end{algorithm}

\section{Types of Meta-paths}
We summarize all the types of meta-paths in Mafengwo and Youshu datasets, and the result is shown in Table \ref{tab:metapath}.
\begin{table}[h]
\caption{The statistics of the metapaths}
\label{tab:metapath}
\centering
\begin{tabular}{|l|l|ll|}
\hline
Scenarios                & Dataset                  & \multicolumn{2}{c|}{Meta-path}                   \\ \hline
\multirow{2}{*}{Group} & \multirow{2}{*}{Mafengwo} & \multicolumn{1}{l|}{Pre-train} & GUG/GUI/IUG/IUI \\
                         &                          & \multicolumn{1}{l|}{Fine-tune} & GIG/GI/IG/IGI   \\ \hline
\multirow{2}{*}{Bundle}  & \multirow{2}{*}{Youshu}  & \multicolumn{1}{l|}{Pre-train} & UIU/UIB/BIU/BIB \\
                         &                          & \multicolumn{1}{l|}{Fine-tune} & UBU/UB/BU/BUB   \\ \hline
\multirow{2}{*}{General} & Mafengwo                 & \multicolumn{2}{c|}{GIG/GI/IG/IGI}               \\
                         & Youshu                   & \multicolumn{2}{c|}{UBU/UB/BU/BUB}               \\\hline
\end{tabular}
\end{table}

\section{Supplementary Experiments}
\label{section:supplementary experiments}
\subsection{Generality of the pre-training Task}
\begin{figure}[h]
	\centering
	\scalebox{0.4}{
	\includegraphics[width=15cm]{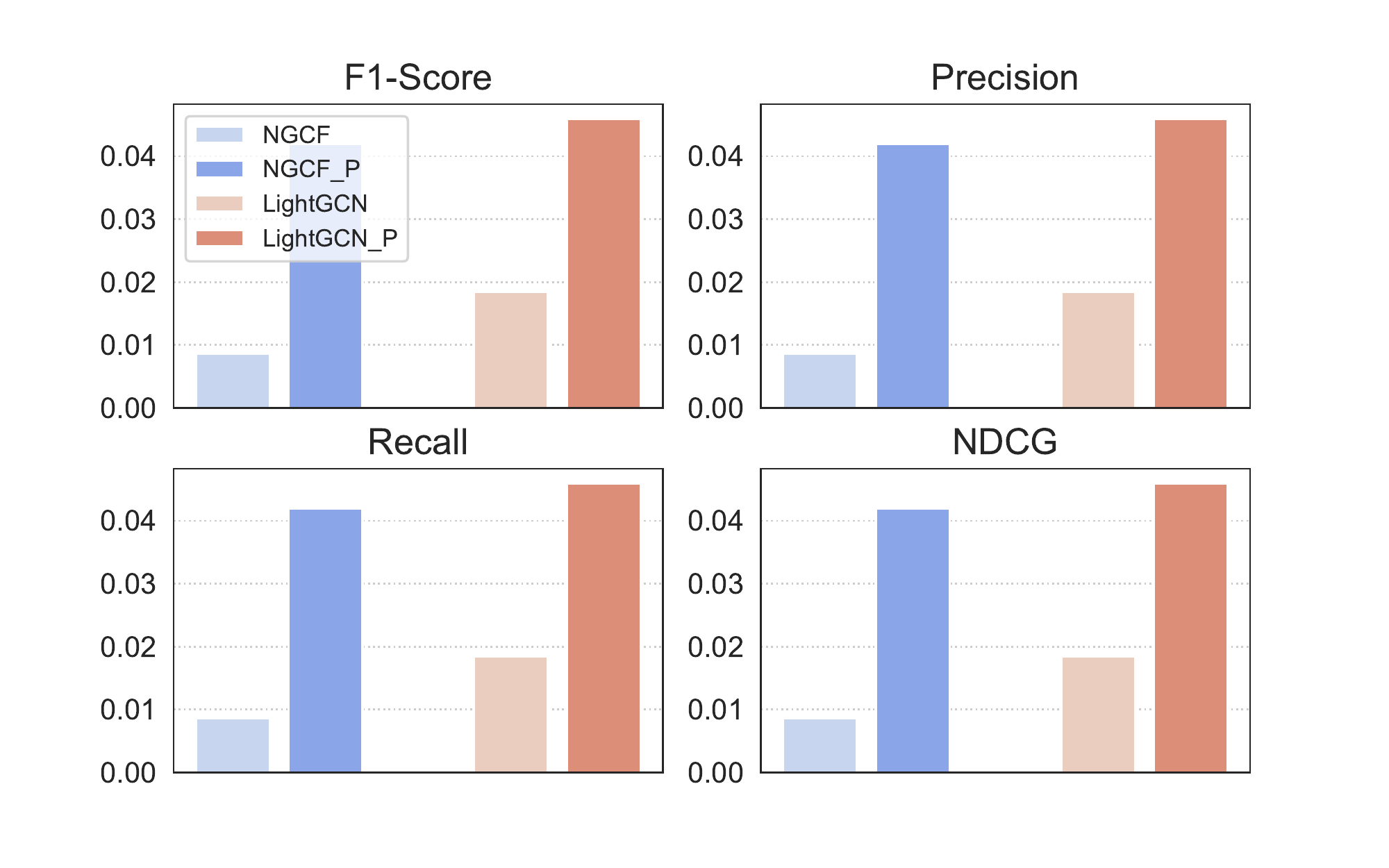}}
	\caption{The performance comparison of the two baselines on the Mafengwo dataset before and after EXTRE\_P pre-training.}
	\label{fig:pre-train}
\end{figure}

Pre-train\&fine-tune models can be trained independently from downstream tasks. Therefore, we perform pre-training on two outstanding models through the EXTRE\_P to verify the generality of our pre-training model. As shown in Fig. \ref{fig:pre-train}, \textit{P} represents models with EXTRE\_P pre-training. After EXTRE\_P pre-training, various indicators of these two baselines have doubled. This result shows that user-item interactions and group-user affiliations are critical for group recommendation. Performing the group recommendation through general recommendation models would waste vastly valuable information. 

\subsection{Hyper-Parameter Sensitivity}
\begin{figure}[h]
	\centering
	\scalebox{0.4}{
	\includegraphics[width=15cm]{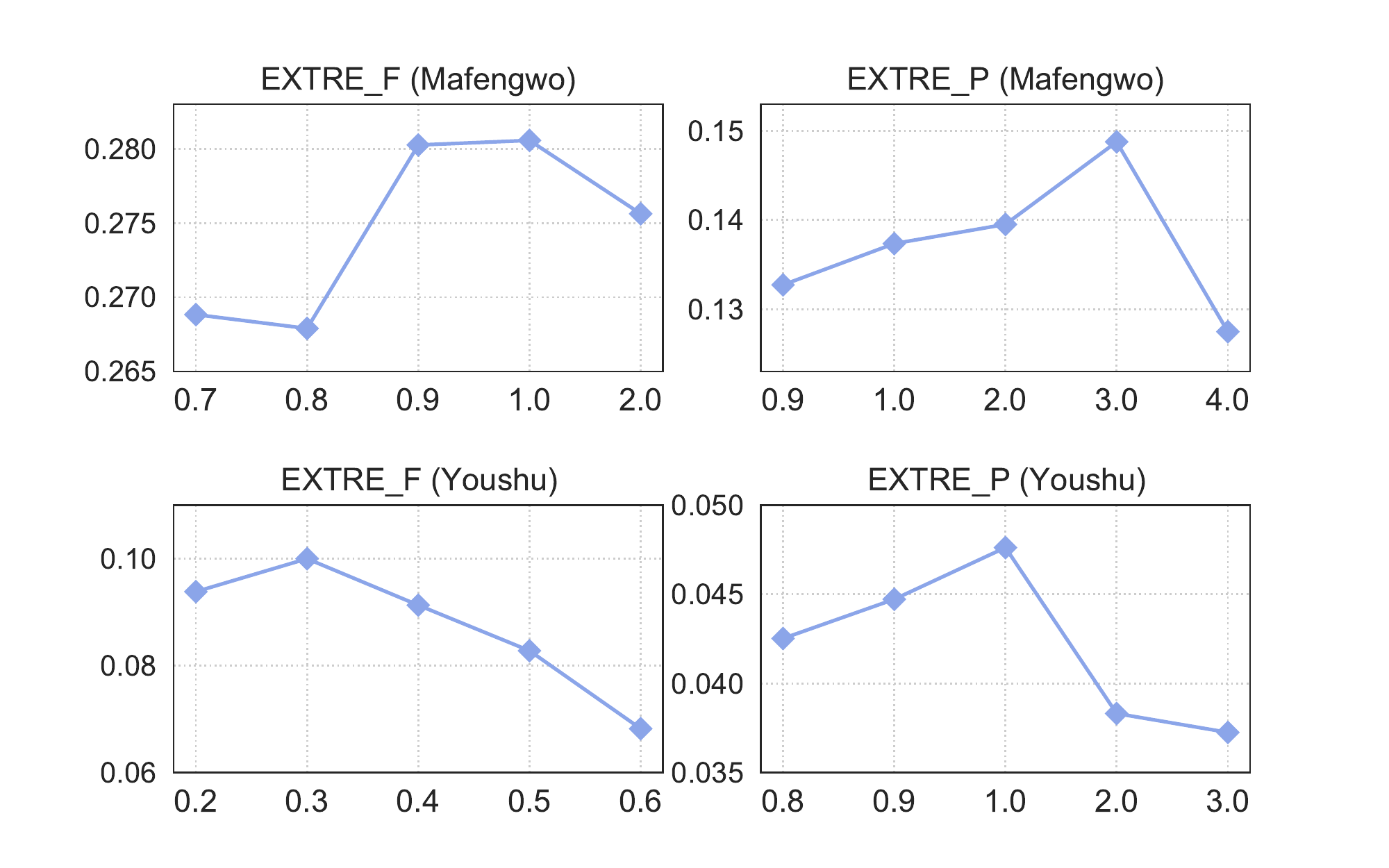}}
	\caption{The effect of different temperatures $\tau$ on the pre-training and fine-tune phases with Recall as the metric}
	\label{fig:t}
\end{figure}

To guide parameter selection, we conduct hyper-parameter sensitivity experiments on EXTRE. Figure \ref{fig:t} shows the effect of different temperature $\tau$ in pre-training and fine-tuning phases of Mafengwo and Youshu datasets. It can be clearly seen that EXTRE is sensitive to the temperature $\tau$. The optimal temperature value in the pre-training stage is higher than that in the fine-tuning stage on these two datasets. Because user-item interactions are noisier than group-item interactions, the model is easily misled by noisy data. Therefore, a higher $\tau$ is required to distinguish dissimilar nodes and learn approximate node preferences. In the fine-tuning stage, group-item interactions can reflect users' actual interests more directly. Our model EXTRE utilizes lower $\tau$ to distinguish similar nodes, purify node embeddings, and learn more fine-grained node preferences.

\clearpage
\end{document}